\documentclass[conference]{IEEEtran}
\IEEEoverridecommandlockouts
\usepackage{cite}
\usepackage{amsmath,amssymb,amsfonts}
\usepackage{algorithmic}
\usepackage{graphicx}
\usepackage{textcomp}
\usepackage{xcolor}
\def\BibTeX{{\rm B\kern-.05em{\sc i\kern-.025em b}\kern-.08em
    T\kern-.1667em\lower.7ex\hbox{E}\kern-.125emX}}

\usepackage{cite}

\usepackage{booktabs}
\usepackage{siunitx} 
\usepackage[dvipsnames]{xcolor}
\usepackage[hidelinks]{hyperref}
\usepackage{pifont}
\usepackage{enumitem}
\usepackage{enumerate}
\usepackage{amsthm,amssymb}
\usepackage[english]{babel}

\usepackage{hyperref}
\usepackage{threeparttable}
\usepackage{mathtools}
\hypersetup{
	colorlinks=true,
	linkcolor=Emerald,
	filecolor=Emerald,
	citecolor = Emerald,
	urlcolor=Emerald
}

\begin{document}

\title{RAS: A Bit-Exact rANS Accelerator For High-Performance Neural Lossless Compression}

\author{
\IEEEauthorblockN{
Yuchao Qin\textsuperscript{1,2,3},
Anjunyi Fan\textsuperscript{1,2},
Bonan Yan\textsuperscript{1,2,3*}
\IEEEauthorblockA{
$^1$Institute for Artificial Intelligence, Peking University, Beijing, China\\
$^2$School of Electronic Engineering and Computer Science, EECS, Peking University, Beijing, China\\
$^3$School of Integrated Circuits, Peking University, Beijing, China\\
Email: bonanyan@pku.edu.cn
}}
}

\maketitle

\begin{abstract}
Data centers handle vast volumes of data that require efficient lossless compression, yet emerging probabilistic models based methods are often computationally slow. To address this, we introduce \emph{RAS}, the \underline{R}ange Asymmetric Numeral System \underline{A}cceleration \underline{S}ystem, a hardware architecture that integrates the rANS algorithm into a lossless compression pipeline and eliminates key bottlenecks. RAS couples an rANS core with a probabilistic generator, storing distributions in BF16 format and converting them once into a fixed-point domain shared by a unified division/modulo datapath. A two-stage rANS update with byte-level re-normalization reduces logic cost and memory traffic, while a \emph{prediction-guided decoding} path speculatively narrows the cumulative distribution function (CDF) search window and safely falls back to maintain bit-exactness. A multi-lane organization scales throughput and enables fine-grained clock gating for efficient scheduling. On image workloads, our RTL-simulated prototype achieves 121.2$\times$ encode and 70.9$\times$ decode speedups over a Python rANS baseline, reducing average decoder binary-search steps from 7.00 to 3.15 (approximately 55\% fewer). When paired with neural probability models, RAS sustains higher compression ratios than classical codecs and outperforms CPU/GPU rANS implementations, offering a practical approach to fast neural lossless compression.
\end{abstract}


\section{Introduction}

The proliferation of data-intensive applications--such as image and video streaming alongside real-time analytics—has intensified the enduring trade-off between compression efficiency and throughput. While lossy compression schemes can reduce bit rates at the expense of distortion~\cite{shannon1948mathematical, Elakkiya2022Comprehensive}, many deployments necessitate bit-exact recovery. This drives the adoption of lossless compression pipelines based on entropy coding, including Huffman coding~\cite{huffman1952method}, arithmetic coding~\cite{CAAC, Brotli}, mix-algorithm~\cite{zstd, Brotli}, standards like JPEG, JPEG-LS, and PNG~\cite{JPEG, JPEG-LS, JPEG-LS_madical}, WebP~\cite{WebP}, as well as modern asymmetric numeral systems (ANS) variants~\cite{rANS}. However, conventional software implementations often fail to achieve both high compression ratios and low latency under real-time and energy constraints in very-large-scale datacenter application scenarios, highlighting a critical area for improvement.

\begin{figure}[t]
    \centering
    \includegraphics[width=0.5\textwidth]{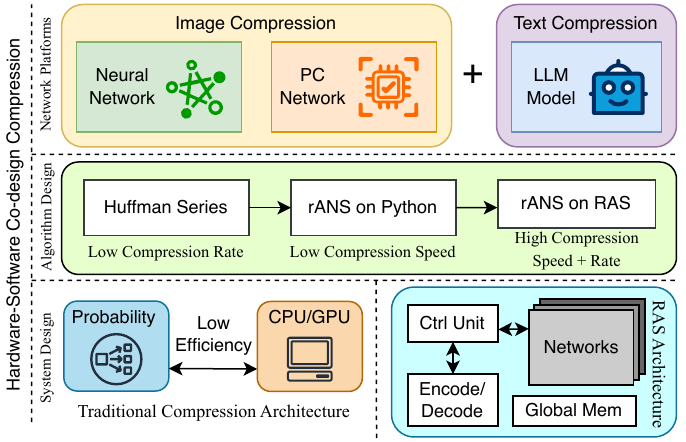}
    \vspace{-1em}
    \caption{Overview of the hardware-software codesign for learned lossless compression. \textit{Modeling:} neural/PC models for images and LLMs for text provide calibrated probabilities. \textit{Algorithmic:} progression from Huffman and software rANS to \textbf{rANS on \textit{RAS}} (this work) increases throughput and compression ratio. \textit{System/Hardware:} RAS integrates a control unit, networks, pipelined encode/decode, and shared global memory--replacing a low-efficiency CPU/GPU + probability generator while preserving bit-exactness.}
    \label{fig:overview_work}
    \vspace{-1.5em}
\end{figure}

To address these challenges, an emerging strategy integrates integer entropy coding with learned probabilistic models. Specifically, the range variant of asymmetric numeral systems (rANS)~\cite{rANS} is highly amenable to hardware implementation and, when paired with context-conditioned distributions from neural or probabilistic predictors, can achieve near-entropy performance. Recent compressors leveraging large models exemplify this paradigm by incorporating autoregressive probabilities into arithmetic or rANS coders, achieving state-of-the-art lossless compression rates for diverse data types such as text, images, audio, and video~\cite{LMCompress2025, CALIC, stream-based-compression, video-compression, Enttsel2024AICAS}. The hardware-software codesign landscape (encompassing modeling techniques, algorithmic selections, and system integration) is summarized in Fig.~\ref{fig:overview_work}.

\begin{figure*}[t]
	\centering
	\includegraphics[width=\textwidth]{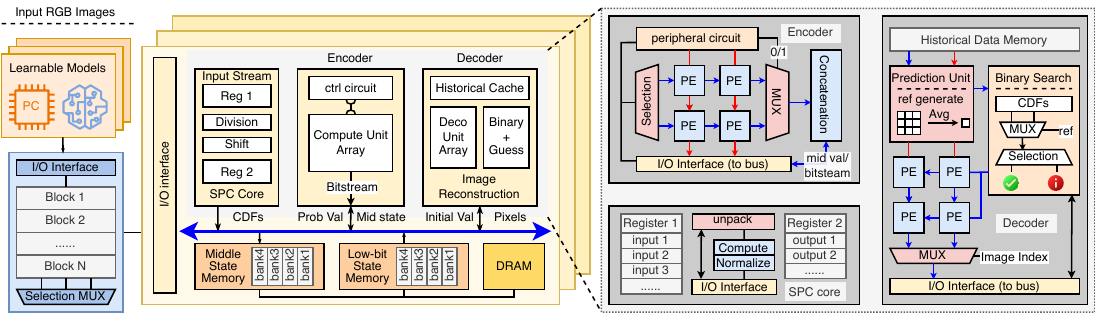}
	\caption{Overall \textit{RAS} architecture. \textbf{Left:} Learnable models (orange block) produce absolute distributions that are stored in Global Memory (blue block); the \textit{SPC} performs a single BF16$\rightarrow$fixed-point conversion with mass correction and streams shared CDF/frequency tables. \textbf{Middle:} The rANS \emph{Encoder} and \emph{Decoder} share a mixed-precision div/mod datapath structure with a two-stage update (parallel quotient/remainder) and byte-level re-normalization; per-lane \emph{Middle-State} and \emph{Low-bit State} memories sustain throughput. \textbf{Right:} A prediction-guided decoding path proposes a trial symbol and verifies it against the CDF, falling back on mismatch—reducing average search while preserving bit-exactness. A simple multi-lane fabric with arbitration and clock gating scales throughput without changing the bitstream.}
	\label{fig:architecture_all}
\end{figure*}

Building on this codesign framework, this paper introduces \emph{RAS}, an rANS-based accelerator design, aiming to maintain determinism and bit-exact recovery while overcoming the key throughput limitations in learned lossless compression. The architecture incorporates three key circuit and architecture design techniques: (a) a unified mixed-precision division/modulus datapath with byte-level re-normalization to minimize logic and memory overhead; (b) a two-stage rANS update that optimizes the balance between arithmetic operations and normalization for sustained high throughput; and (c) a \emph{prediction-guided decoding} technique that speculatively reduces the cumulative distribution function (CDF) search space with fallback mechanisms to preserve bitstream integrity.
With RTL-based simulation experiments on image workloads, \textit{RAS} achieves 121.2$\times$ encode and 70.9$\times$ decode speedups over a Python rANS baseline, and reduces average decoder binary-search steps from 7.00 to 3.15 ($\approx55\%$ fewer).

\section{Preliminaries}\label{sec:prelim}

\subsection{rANS Algorithm Acceleration}

rANS approaches Shannon’s limit~\cite{shannon1948mathematical} by updating a single integer state using table-indexed operations with power-of-two re-normalization, making it well suited to hardware compare to other entropy coding. In modern compressors~\cite{GPU-entropy-codec, Dual-attention-compression}, rANS is paired with learned priors—e.g., probabilistic circuits (PC)~\cite{LiuPC} or compact neural networks—that emit context-conditioned symbol distributions subsequently quantized into frequencies or CDFs~\cite{LiuPC}. Because the rANS pipeline consists of two tightly coupled kernels (state update in encoding and state-to-symbol inversion in decoding), both directions must be accelerated without breaking bit-exactness. Prior work typically optimizes only one side, or omits a hardware–software co-design that simultaneously reduces table-storage pressure, arithmetic cost, and data movement~\cite{Recoil, bi-rANS, fpga-rans}. In contrast, our approach targets both encode and decode with shared mixed-precision tables, balanced div/mod datapaths, and CDF-aware decoding (Sec.~\ref{sec:method}), preserving determinism while raising throughput and energy efficiency.

\subsection{Data Compression on Modern Processors}

Recent advances aim to improve both compression ratio and speed. To raise throughput, dedicated on-chip compression engines~\cite{FLAC, Deepu2016ECG, EIT} avoid host overheads and exploit fine-grained parallelism. To raise efficiency, entropy coding has become central in contemporary pipelines, routinely outperforming traditional schemes when supplied with calibrated probabilities. Within these pipelines, learnable probability generators based on traditional neural networks~\cite{Wang2018JPEGNN, MetaRespiratory2024} or PCs refine per-symbol distributions that feed a downstream coder. However, few hardware systems tightly couple such predictors with the entropy core while controlling memory footprint and interconnect traffic. Our design explicitly pairs a PC-based (or compact NN) probability module with rANS via a shared fixed-point interface and format-preserving mass correction, enabling higher compression ratios and lower latency without altering the bitstream; implementation details appear in Sec.~\ref{sec:method}.

\section{System Overview}\label{sec:overall}

Figure~\ref{fig:architecture_all} illustrates the RAS datapath and its interfaces. A model-agnostic \emph{neural compression module} (e.g., PC or compact NN) generates \emph{absolute} symbol distributions, which are stored in \emph{Global Memory} using BF16 precision. The \emph{Streaming Prefetch Converter (SPC)} then converts these distributions once into fixed-point CDF or frequency tables, aligned with the rANS radix, and streams them via a backpressure-aware on-chip \emph{bus}. The rANS fabric comprises an \emph{Encoder} and a \emph{Decoder} that share a mixed-precision division and modulo datapath. Each engine incorporates local control logic and \emph{Middle-State (MS) Memory} to sustain line-rate operation while preserving determinism and bit-exactness.

\begin{figure}[t]
    \centering
    \includegraphics[width=0.5\textwidth]{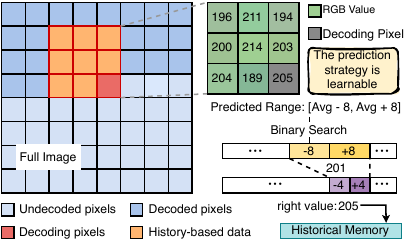}
    \caption{Prediction-guided rANS decoding: the neighborhood average (201) defines a window $[\mathrm{Avg}-8,\mathrm{Avg}+8]$; a dichotomous refinement ($\pm8 \rightarrow \pm4$) resolves the symbol, yielding the correct value 205.}
    \label{fig:prediction}
\end{figure}

Building on this architecture, the Encoder employs a stationary dataflow where the rANS state and symbols remain resident, while probability entries stream through the shared arithmetic unit. The update process involves a two-stage pipeline: parallel quotient and remainder paths followed by byte-level re-normalization, which maintains high pipeline occupancy and amortizes conversion overhead off the critical path. The Decoder mirrors this structure but focuses on state-to-symbol recovery. To enhance efficiency, a baseline gated binary search over the CDF is accelerated via \emph{prediction-guided decoding} (shown in Fig.\,\ref{fig:prediction}): a lightweight trial symbol and narrow window are derived from local context, promptly verified against the CDF, and either committed or rolled back with bounded cost, reducing average search depth without modifying bitstream semantics.

Furthermore, parallelism is implemented through multiple independent lanes that operate behind the shared \emph{SPC} and \emph{Global Memory} complex. Each lane autonomously issues prefetches, maintains private \emph{MS Memory}, and arbitrates for \emph{BUS} access via a simple credit scheme; stalled lanes relinquish bandwidth to ready lanes, enhancing utilization and enabling lane-level clock gating. Crucially, since probability generation and rANS formatting remain unmodified, the \emph{RAS} architecture ensures full interoperability with existing learned models and software decoders.

\section{Design Techniques}\label{sec:method}

\subsection{Mixed-Precision Probability Module}

We propose a mixed-precision processing method for rANS encoding and decoding, which optimizes storage and computational efficiency. Conditional and cumulative probability tables are stored in BF16 format, while the rANS state is retained as a 32-bit unsigned integer. A mixed-precision unit performs a single conversion of BF16 inputs into a shared fixed-point domain, with fractional precision aligned to the re-normalization radix \(2^{n}\). Given calibrated probabilities \(\{p_x\}\), integer frequencies are computed as \(f(x) = \max(1, \mathrm{round}(p_x 2^{n}))\), followed by a brief mass-correction pass to ensure \(\sum_x f(x) = 2^{n}\) and that the CDF \(C(x) = \sum_{y<x} f(y)\) remains strictly monotonic. The CDF is cached for reuse by both encoder and decoder.
Subsequent division and remainder operations are executed entirely in the fixed-point domain, avoiding repeated type conversions on the critical computational path. Exponent alignment, incorporating a small guard margin, bounds conversion errors to within one unit in the last place (ULP) of the fixed-point representation. A pipelined divider generates the quotient, while a matched path computes the remainder; both results are power-of-two normalized and forwarded as 32-bit values. Deterministic, monotone rounding (using floor for the quotient) preserves rANS invariants and guarantees bit-exact encoder-decoder consistency across lanes and runs.
This design halves table storage compared to single-precision approaches, reduces off-chip bandwidth demands, and simplifies timing closure at high operating frequencies.

\subsection{Encoding Pipeline}
Let $s_{i-1}$ be the rANS state before symbol $x_i$, $f(x_i)$ its quantized frequency, and $C(x_i)$ the CDF value immediately preceding $x_i$. With re-normalization radix $R=2^{n}$, the range-ANS update can be written as
\begin{equation}
\label{eq:rans_update}
s_i \;=\; \Big\lfloor \frac{s_{i-1}}{f(x_i)} \Big\rfloor \cdot R \;+\; \big(s_{i-1} \bmod f(x_i)\big) \;+\; C(x_i),
\end{equation}
followed by re-normalization to the canonical unsigned range. We exploit the algebraic separability of the quotient and remainder paths by computing
$a_1=\big\lfloor s_{i-1}/f(x_i)\big\rfloor \cdot R$ and
$a_2=(s_{i-1} \bmod f(x_i)) + C(x_i)$
in parallel using the shared fixed-point probabilities, then forming $s_i=a_1+a_2$ before byte-level re-normalization. re-normalization emits one byte per step while maintaining $s\!\in\![L,RL)$ for a fixed lower bound $L$, ensuring identical coder/decoder state evolution. Common normalization and exponent handling are performed once and fanned out to both arithmetic paths, reducing redundant logic and balancing pipeline stages. In practice, this two-stage organization improves steady-state throughput (one symbol per cycle after fill) while preserving determinism and exactness.

\subsection{Speculative Prediction for rANS Decoding}

\begin{figure*}[h]
    \centering
    \includegraphics[width=\textwidth]{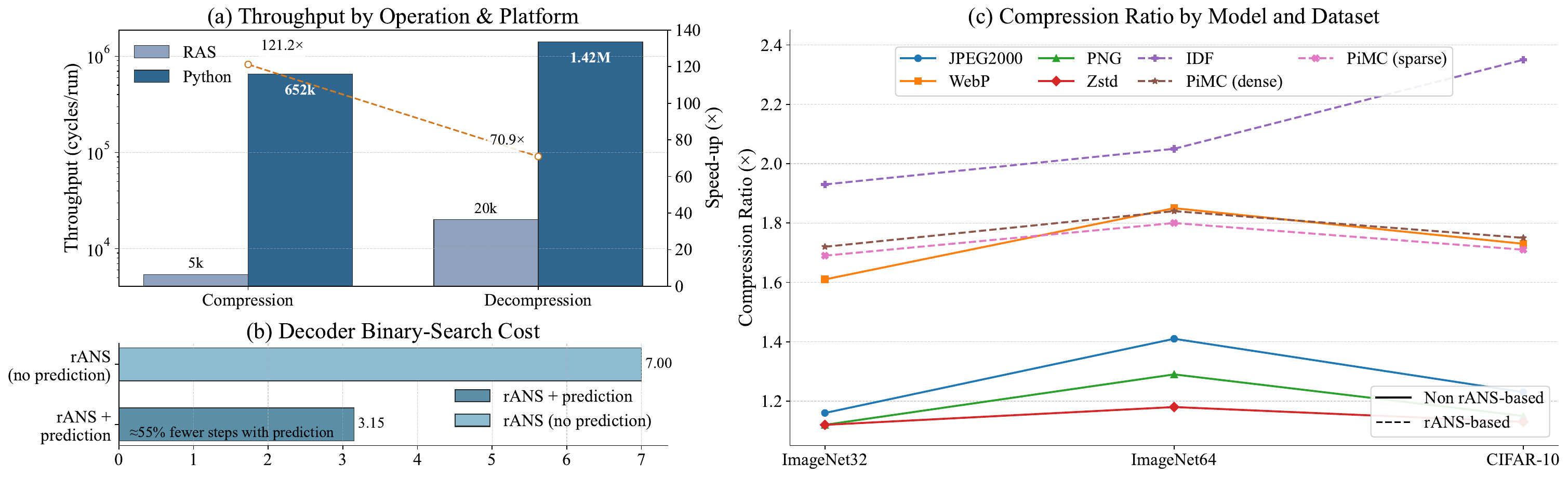} 
    \vspace{-1.5em}
    \caption{Design exploration results (simulated): (a) Cycle-normalized compute cost (cycles/run; lower is better) for compression and decompression, comparing RAS with a Python rANS baseline; annotations show speed-ups of $121.2\times$ (encode) and $70.9\times$ (decode). (b) Decoder binary-search cost: average steps per symbol drop from $7.00$ to $3.15$ with prediction ($\approx55\%$ fewer). (c) Compression ratio on ImageNet32/64 and CIFAR-10 for classical codecs (solid) and neural rANS-based models (dashed); neural models with rANS algorithm (IDF, PiMC) achieve higher ratios.}
    \label{fig:data}
        \vspace{-1.5em}
\end{figure*}

Decoder latency is often dominated by interactions with probability tables and the binary search over the CDF required to map the state back to a symbol. Inspired by speculate-and-verify control and branch prediction in CPU~\cite{TAGE, HW-codesign-prediction}, we introduce a \emph{decoder-side} prediction path that proposes a trial symbol from lightweight context, advances the rANS state as if the proposal were correct, and immediately verifies this advancement against the reference CDF. On a match, the decoder commits and bypasses the full search; on a mismatch, a short, bounded restore returns to the precise pre-speculation state and the conventional lookup proceeds. Because speculation never alters the bitstream or frequency tables, rANS invariants are maintained and worst-case latency remains identical to the baseline.

The predictor emits an anchor $\mu$ and a tolerance $\Delta$, defining a search bracket $[\mu-\Delta,\mu+\Delta]$ over the symbol alphabet. The controller performs a window-gated binary check centered at $\mu$ and refines only when necessary, thereby reducing the effective domain from $|\Sigma|$ to $|R|=2\Delta+1$. The expected search depth drops from $\log_2|\Sigma|$ to $\log_2|R|+\nu$, where $\nu\!\ll\!1$ captures a small amortized verification overhead hidden by the pipeline. If the true symbol lies outside the bracket, the controller expands the window and falls back to the baseline search with a bounded penalty. Unlike encoder-side predictors in PNG/JPEG-LS that reduce residual entropy, our mechanism targets the decoder inner loop to prune CDF work. We deliberately employ hardware-cheap statistics (e.g., neighbor averages with last-value/zero fallback and simple pattern cues) to keep latency fixed and area low while materially reducing table probes and memory traffic; more expressive fixed-point predictors can be plugged in without changing the interface.

\section{Evaluation}\label{sec:evaluation}
\textbf{Simulated Design Exploration:} In this work, we perform comprehensive RTL simulation along two axes: (i) end-to-end encode/decode throughput and (ii) compression ratio on standard image benchmarks. The rANS core is agnostic to the probability generator and interoperates with learned priors, including IDF~\cite{IDF} and PiMC. To isolate the coder, the probability module first produces all required distributions and writes them to shared memory in BF16; the rANS engines then fetch from this cache during encode/decode. The hardware is implemented in SystemVerilog; cycle counts are obtained from RTL simulation (Icarus Verilog). For software comparison, we employ a Python rANS implementation on an Apple M4 and normalize by \emph{cycles} rather than wall time; when a frequency is required, we conservatively take 2.9\,GHz~\cite{appleM4}. Both systems use the same symbolization and CDFs, so the bitstreams are identical.

\textbf{System Prototype:} With the proposed RAS codesign framework, the lossless compression prototype system is demonstrated at \url{https://vimeo.com/1106157880}.

\subsection{Speed}
Fig.\,\ref{fig:data}(a) reports cycle-normalized compute cost (lower is better). RAS achieves $121.2\times$ speedup for encoding and $70.9\times$ for decoding over the Python baseline. Measurements include the full coder kernels—division/modulus, CDF access, byte-level re-normalization, state moves, and (for decode) state-to-symbol search and verification—and exclude host I/O, dataset parsing, and probability generation. The gains arise from three effects: (i) the two-stage update that balances arithmetic with re-normalization and sustains one-symbol-per-cycle throughput after pipeline fill; (ii) single-pass BF16$\!\rightarrow\!$fixed-point conversion with mass-corrected frequencies, which removes repeated casts from the critical path and reduces bus traffic; and (iii) shared tables plus local MS memories, which improve locality and reduce stalls. Decoder throughput is typically constrained by CDF lookups and the state-to-symbol search. As isolated in Fig.~\ref{fig:data}\,(b), prediction-guided decoding lowers the average binary-search depth from 7.00 to 3.15 steps ($\approx55\%$ fewer). Because each step triggers a probability-table access, the reduction directly translates into lower latency and memory traffic while preserving worst-case behavior.

\subsection{Compression Ratio}
Compression ratios on ImageNet32/64~\cite{imagenet} and CIFAR-10~\cite{cifar-10} are shown in Fig.~\ref{fig:data}\,(c)
\footnote{Throughput in MB/s can be derived from processed bytes, cycles, and clock frequency; we use cycles to avoid cross-platform artifacts.}. We report $\mathrm{CR}=\text{original bytes}/\text{compressed bytes}$ (higher is better). Classical lossless codecs (JPEG2000~\cite{JPEG2000}, WebP, PNG, Zstd; solid curves) improve with increased spatial context but remain below rANS-based neural models (IDF, PiMC dense/sparse; dashed curves) across datasets. RAS is format-preserving: the probability tables, CDF, and symbolization are identical to the software pipeline, and the BF16$\rightarrow$fixed-point conversion uses a deterministic mass-correction that enforces $\sum_x f(x)=2^n$. Consequently, RAS reproduces the exact bitstreams of the reference implementation while accelerating the coding path, so the compression ratios in Fig.~\ref{fig:data}\,(c) transfer unchanged.

\section{Conclusion}
We introduce \emph{RAS}, an accelerator that integrates an rANS core with a mixed-precision probability path, a two-stage update pipeline, and a prediction-guided decoder to optimize CDF searches while preserving bit-exactness. The \emph{predict-and-verify} approach generalizes to ANS variants like tANS and iANS, pruning decoder loops without altering bitstreams, and can enhance arithmetic decoding by biasing interval tests toward probable outcomes.
Future work will (i) measure PPA on silicon and integrate on-chip probability generators, and (ii) explore lightweight ML predictors~\cite{MACRO-CNN-CPU-branch-prediction, ISCA-ML-predictor} that raise accuracy while preserving exactness. Overall, RAS offers a practical, extensible path to high-throughput, bit-exact neural lossless compression using rANS algorithm.

\bibliographystyle{IEEEtran}
\bibliography{refs}

\begin{thebibliography}{10}
\providecommand{\url}[1]{#1}
\csname url@samestyle\endcsname
\providecommand{\newblock}{\relax}
\providecommand{\bibinfo}[2]{#2}
\providecommand{\BIBentrySTDinterwordspacing}{\spaceskip=0pt\relax}
\providecommand{\BIBentryALTinterwordstretchfactor}{4}
\providecommand{\BIBentryALTinterwordspacing}{\spaceskip=\fontdimen2\font plus
\BIBentryALTinterwordstretchfactor\fontdimen3\font minus
  \fontdimen4\font\relax}
\providecommand{\BIBforeignlanguage}[2]{{%
\expandafter\ifx\csname l@#1\endcsname\relax
\typeout{** WARNING: IEEEtran.bst: No hyphenation pattern has been}%
\typeout{** loaded for the language `#1'. Using the pattern for}%
\typeout{** the default language instead.}%
\else
\language=\csname l@#1\endcsname
\fi
#2}}
\providecommand{\BIBdecl}{\relax}
\BIBdecl

\bibitem{shannon1948mathematical}
C.~E. Shannon, ``A mathematical theory of communication,'' \emph{Bell System
  Technical Journal}, vol.~27, no. 3–4, pp. 379--423, 623--656, 1948.

\bibitem{Elakkiya2022Comprehensive}
\BIBentryALTinterwordspacing
S.~Elakkiya and K.~S. Thivya, ``Comprehensive review on lossy and lossless
  compression techniques,'' \emph{Journal of The Institution of Engineers
  (India): Series B}, vol. 103, no.~3, pp. 1003--1012, Jun. 2022. [Online].
  Available: \url{https://doi.org/10.1007/s40031-021-00686-3}
\BIBentrySTDinterwordspacing

\bibitem{huffman1952method}
D.~A. Huffman, ``A method for the construction of minimum-redundancy codes,''
  \emph{Proceedings of the IRE}, vol.~40, no.~9, pp. 1098--1101, 1952.

\bibitem{CAAC}
J.-J. Ding, I.-H. Wang, and H.-Y. Chen, ``Improved efficiency on adaptive
  arithmetic coding for data compression using range-adjusting scheme,
  increasingly adjusting step, and mutual-learning scheme,'' \emph{IEEE
  Transactions on Circuits and Systems for Video Technology}, vol.~28, no.~12,
  pp. 3412--3423, 2018.

\bibitem{Brotli}
\BIBentryALTinterwordspacing
J.~Alakuijala, A.~Farruggia, P.~Ferragina, E.~Kliuchnikov, R.~Obryk,
  Z.~Szabadka, and L.~Vandevenne, ``Brotli: A general-purpose data
  compressor,'' \emph{ACM Trans. Inf. Syst.}, vol.~37, no.~1, Dec. 2018.
  [Online]. Available: \url{https://doi.org/10.1145/3231935}
\BIBentrySTDinterwordspacing

\bibitem{zstd}
\BIBentryALTinterwordspacing
Y.~Collet and M.~Kucherawy, ``{Zstandard Compression and the 'application/zstd'
  Media Type},'' RFC 8878, Feb. 2021. [Online]. Available:
  \url{https://www.rfc-editor.org/info/rfc8878}
\BIBentrySTDinterwordspacing

\bibitem{JPEG}
\BIBentryALTinterwordspacing
G.~K. Wallace, ``The jpeg still picture compression standard,'' \emph{Commun.
  ACM}, vol.~34, no.~4, p. 30–44, Apr. 1991. [Online]. Available:
  \url{https://doi.org/10.1145/103085.103089}
\BIBentrySTDinterwordspacing

\bibitem{JPEG-LS}
\BIBentryALTinterwordspacing
J.~Hua, H.~Xu, Y.~Du, and L.~Du, ``Improved jpeg lossless compression for
  compression of intermediate layers in neural networks based on
  compute-in-memory,'' \emph{Electronics}, vol.~13, no.~19, 2024. [Online].
  Available: \url{https://www.mdpi.com/2079-9292/13/19/3872}
\BIBentrySTDinterwordspacing

\bibitem{JPEG-LS_madical}
S.-G. Miaou, F.-S. Ke, and S.-C. Chen, ``A lossless compression method for
  medical image sequences using jpeg-ls and interframe coding,'' \emph{IEEE
  Transactions on Information Technology in Biomedicine}, vol.~13, no.~5, pp.
  818--821, 2009.

\bibitem{WebP}
\BIBentryALTinterwordspacing
J.~Zern, P.~Massimino, and J.~Alakuijala, ``Webp image format,'' RFC Editor,
  Request for Comments 9649, Nov. 2024. [Online]. Available:
  \url{https://www.rfc-editor.org/rfc/rfc9649}
\BIBentrySTDinterwordspacing

\bibitem{rANS}
J.~Duda, ``Asymmetric numeral systems as close to capacity low state entropy
  coders,'' 11 2013.

\bibitem{LMCompress2025}
\BIBentryALTinterwordspacing
Z.~Li, C.~Huang, X.~Wang, H.~Hu, C.~Wyeth, D.~Bu, Q.~Yu, W.~Gao, X.~Liu, and
  M.~Li, ``Lossless data compression by large models,'' \emph{Nature Machine
  Intelligence}, vol.~7, no.~7, pp. 794--799, May 2025. [Online]. Available:
  \url{https://doi.org/10.1038/s42256-025-01033-7}
\BIBentrySTDinterwordspacing

\bibitem{CALIC}
J.-M. Chang, J.-J. Ding, and H.-S. Lin, ``Adaptive prediction, context
  modeling, and entropy coding methods for calic lossless image compression,''
  in \emph{2019 IEEE Asia Pacific Conference on Circuits and Systems (APCCAS)},
  2019, pp. 349--352.

\bibitem{stream-based-compression}
S.~Yamagiwa and S.~Kuwabara, ``Autonomous parameter adjustment method for
  lossless data compression on adaptive stream-based entropy coding,''
  \emph{IEEE Access}, vol.~8, pp. 186\,890--186\,903, 2020.

\bibitem{video-compression}
D.~Ammous, A.~Kessentini, N.~Khlif, F.~Kammoun, and N.~Masmoudi, ``Evaluation
  of the improvement in hierarchical lossless video compression,'' in
  \emph{2023 9th International Conference on Control, Decision and Information
  Technologies (CoDIT)}, 2023, pp. 01--06.

\bibitem{Enttsel2024AICAS}
A.~Enttsel, A.~Marchioni, G.~Setti, M.~Mangia, and R.~Rovatti, ``Enhancing
  anomaly detection with entropy regularization in autoencoder-based
  lightweight compression,'' in \emph{2024 IEEE 6th International Conference on
  AI Circuits and Systems (AICAS)}, 2024, pp. 273--277.

\bibitem{GPU-entropy-codec}
H.~Akutsu, T.~Naruko, and A.~Suzuki, ``Gpu-intensive fast entropy coding
  framework for neural image compression,'' in \emph{2021 International
  Conference on Visual Communications and Image Processing (VCIP)}, 2021, pp.
  1--5.

\bibitem{Dual-attention-compression}
W.~Tianwen, ``Dual attention entropy model for efficient neural image
  compression,'' in \emph{2024 21st International Computer Conference on
  Wavelet Active Media Technology and Information Processing (ICCWAMTIP)},
  2024, pp. 1--5.

\bibitem{LiuPC}
\BIBentryALTinterwordspacing
A.~Liu, S.~Mandt, and G.~Van~den Broeck, ``Lossless compression with
  probabilistic circuits,'' in \emph{Proceedings of the International
  Conference on Learning Representations (ICLR)}, apr 2022. [Online].
  Available: \url{http://starai.cs.ucla.edu/papers/LiuICLR22.pdf}
\BIBentrySTDinterwordspacing

\bibitem{Recoil}
\BIBentryALTinterwordspacing
F.~Lin, K.~Arunruangsirilert, H.~Sun, and J.~Katto, ``Recoil: Parallel rans
  decoding with decoder-adaptive scalability,'' in \emph{Proceedings of the
  52nd International Conference on Parallel Processing}, ser. ICPP '23.\hskip
  1em plus 0.5em minus 0.4em\relax New York, NY, USA: Association for Computing
  Machinery, 2023, p. 31–40. [Online]. Available:
  \url{https://doi.org/10.1145/3605573.3605588}
\BIBentrySTDinterwordspacing

\bibitem{bi-rANS}
E.~Belyaev and K.~Liu, ``An adaptive binary rans with probability estimation in
  reverse order,'' \emph{IEEE Signal Processing Letters}, vol.~30, pp.
  1487--1491, 2023.

\bibitem{fpga-rans}
M.~Pastuła, P.~Russek, and K.~Wiatr, ``Low-cost ans encoder for lossless data
  compression in fpgas,'' \emph{International Journal of Electronics and
  Telecommunications}, pp. 219--219, 03 2024.

\bibitem{FLAC}
\BIBentryALTinterwordspacing
M.~van Beurden and A.~Weaver, ``{Free Lossless Audio Codec (FLAC)},'' RFC 9639,
  Dec. 2024. [Online]. Available: \url{https://www.rfc-editor.org/info/rfc9639}
\BIBentrySTDinterwordspacing

\bibitem{Deepu2016ECG}
C.~J. Deepu, X.~Zhang, C.~H. Heng, and Y.~Lian, ``A 3-lead ecg-on-chip with qrs
  detection and lossless compression for wireless sensors,'' \emph{IEEE
  Transactions on Circuits and Systems II: Express Briefs}, vol.~63, no.~12,
  pp. 1151--1155, 2016.

\bibitem{EIT}
B.~Liu, C.-H. Heng, G.~Wang, and Y.~Lian, ``On-chip data compression scheme for
  lung eit signal acquisition and recovery,'' in \emph{2018 IEEE 23rd
  International Conference on Digital Signal Processing (DSP)}, 2018, pp. 1--5.

\bibitem{Wang2018JPEGNN}
Z.~Wang, S.~Yin, F.~Tu, L.~Liu, and S.~Wei, ``An energy efficient jpeg encoder
  with neural network based approximation and near-threshold computing,'' in
  \emph{2018 IEEE International Symposium on Circuits and Systems (ISCAS)},
  2018, pp. 1--5.

\bibitem{MetaRespiratory2024}
Q.~Zhang, C.~Chen, S.~Yuan, J.~Zhang, J.~Yuan, H.~Huang, Y.~Zhang, R.~Pan,
  X.~Jiang, J.~Zhao, Y.~Li, Y.~Yin, L.~Zhao, G.~Wang, and Y.~Lian, ``Meta: Data
  compression and event detection grand challenge 2024 with sprsound dataset,''
  \emph{IEEE Data Descriptions}, vol.~1, pp. 122--130, 2024.

\bibitem{TAGE}
A.~Seznec and P.~Michaud, ``A case for (partially) tagged geometric history
  length branch prediction,'' \emph{Journal of Instruction-level Parallelism -
  JILP}, vol.~8, 02 2006.

\bibitem{HW-codesign-prediction}
T.~A. Khan, M.~Ugur, K.~Nathella, D.~Sunwoo, H.~Litz, D.~A. Jiménez, and
  B.~Kasikci, ``Whisper: Profile-guided branch misprediction elimination for
  data center applications,'' in \emph{2022 55th IEEE/ACM International
  Symposium on Microarchitecture (MICRO)}, 2022, pp. 19--34.

\bibitem{IDF}
E.~Hoogeboom, J.~W. Peters, R.~van~den Berg, and M.~Welling, \emph{Integer
  discrete flows and lossless compression}.\hskip 1em plus 0.5em minus
  0.4em\relax Red Hook, NY, USA: Curran Associates Inc., 2019.

\bibitem{appleM4}
Notebookcheck, ``Apple m4 (10 cores) processor – benchmarks and specs,''
  \url{https://www.notebookcheck.net/Apple-M4-10-cores-Processor-Benchmarks-and-Specs.835975.0.html},
  2024, accessed: 2025-10-07.

\bibitem{imagenet}
J.~Deng, W.~Dong, R.~Socher, L.-J. Li, K.~Li, and L.~Fei-Fei, ``Imagenet: A
  large-scale hierarchical image database,'' in \emph{2009 IEEE Conference on
  Computer Vision and Pattern Recognition}, 2009, pp. 248--255.

\bibitem{cifar-10}
\BIBentryALTinterwordspacing
A.~Krizhevsky, ``Learning multiple layers of features from tiny images,''
  University of Toronto, Tech. Rep. TR-2009, 2009, accessed: 2025-10-07.
  [Online]. Available:
  \url{https://www.cs.toronto.edu/~kriz/learning-features-2009-TR.pdf}
\BIBentrySTDinterwordspacing

\bibitem{JPEG2000}
D.~Taubman and M.~Marcellin, ``Jpeg2000: standard for interactive imaging,''
  \emph{Proceedings of the IEEE}, vol.~90, no.~8, pp. 1336--1357, 2002.

\bibitem{MACRO-CNN-CPU-branch-prediction}
S.~Zangeneh, S.~Pruett, S.~Lym, and Y.~N. Patt, ``Branchnet: A convolutional
  neural network to predict hard-to-predict branches,'' in \emph{2020 53rd
  Annual IEEE/ACM International Symposium on Microarchitecture (MICRO)}, 2020,
  pp. 118--130.

\bibitem{ISCA-ML-predictor}
E.~Garza, S.~Mirbagher-Ajorpaz, T.~A. Khan, and D.~A. Jiménez, ``Bit-level
  perceptron prediction for indirect branches,'' in \emph{2019 ACM/IEEE 46th
  Annual International Symposium on Computer Architecture (ISCA)}, 2019, pp.
  27--38.

\end{thebibliography}

\end{document}